\documentclass{article}
\usepackage{graphicx,subfigure,a4wide}
\usepackage{xspace}

\newcommand{\Bc}{\ensuremath{B_c}\xspace}
\newcommand{\BcJpsipi}{\ensuremath{\Bc\rightarrow\Jpsi\pi^+}\xspace}

\newcommand{\Bplus}{\ensuremath{B^+}\xspace}
\newcommand{\BplusJpsiK}{\ensuremath{\Bplus\rightarrow\Jpsi K^+}\xspace}
\newcommand{\BsDsonemunu}{\ensuremath{\Bs\rightarrow\DsSSone \mu^+\nu X}\xspace}
\newcommand{\Bs}{\ensuremath{B_s}\xspace}
\newcommand{\BS}{\ensuremath{B^*}\xspace}

\newcommand{\BSS}{\ensuremath{B^{**0}}\xspace}

\newcommand{\BSSone}{\ensuremath{B_1^0}\xspace}

\newcommand{\BsSS}{\ensuremath{B_s^{**0}}\xspace}

\newcommand{\BsSSone}{\ensuremath{B_{s1}^0}\xspace}

\newcommand{\BsSStwo}{\ensuremath{B_{s2}^{*0}}\xspace}

\newcommand{\BSStwo}{\ensuremath{B^{*0}_2}\xspace}

\newcommand{\DObar}{\ensuremath{\overline{D}^0}\xspace}
\newcommand{\DSplus}{\ensuremath{D^{*+}}\xspace}

\newcommand{\DsSSone}{\ensuremath{D_{s1}^-(2536)}\xspace}
\newcommand{\gevcc}{\ensuremath{\mathrm{GeV}/c^2}\xspace}
\newcommand{\invfb}{\ensuremath{\mathrm{fb^{-1}}}\xspace}
\newcommand{\invpb}{\ensuremath{\mathrm{pb^{-1}}}\xspace}
\newcommand{\Jpsi}{\ensuremath{J\!/\!\psi}\xspace}
\newcommand{\Lambdab}{\ensuremath{\Lambda_b^0}\xspace}
\newcommand{\mevcc}{\ensuremath{\mathrm{MeV}/c^2}\xspace}
\newcommand{\mev}{\ensuremath{\mathrm{MeV}}\xspace}
\newcommand{\Omegab}{\ensuremath{\Omega_b}\xspace}
\newcommand{\ps}{\ensuremath{\mathrm{ps}}\xspace}
\newcommand{\Sigmab}{\ensuremath{\Sigma_b^\pm}\xspace}
\newcommand{\SigmabS}{\ensuremath{\Sigma_b^{*\pm}}\xspace}

\newcommand{\stat}{\ensuremath{\mathrm{(stat)}}\xspace}
\newcommand{\sys}{\ensuremath{\mathrm{(sys)}}\xspace}
\newcommand{\XibDLambda}{\ensuremath{\Xib\rightarrow D\Lambda}\xspace}
\newcommand{\Xib}{\ensuremath{\Xi_b}\xspace}
\newcommand{\XibJpsiXi}{\ensuremath{\Xib\rightarrow\Jpsi\Xi}\xspace}
\newcommand{\XibLambdacKpi}{\ensuremath{\Xib\rightarrow \Lambdac K\pi}\xspace}
\newcommand{\Xibminus}{\ensuremath{\Xi_b^-}\xspace}
\newcommand{\XibminusJpsiXi}{\ensuremath{\Xibminus\rightarrow\Jpsi\Xi^-}\xspace}
\newcommand{\XibminusXicpi}{\ensuremath{\Xib^-\rightarrow\Xic^0\pi^-}\xspace}
\newcommand{\XibXicpi}{\ensuremath{\Xib\rightarrow\Xic\pi}\xspace}
\newcommand{\Xibzero}{\ensuremath{\Xi_b^0}\xspace}
\newcommand{\Xic}{\ensuremath{\Xi_c}\xspace}
\newcommand{\Lambdac}{\ensuremath{\Lambda_c}\xspace}

\begin{document}
%
%
\title{ 
D AND B MESON SPECTROSCOPY, NEW STATES, BARYONS AT THE
TEVATRON
}
\author{
Michal Kreps on behalf of the CDF and D\O{} Collaborations\\
{\em Universit\"at Karlsruhe (TH), Postfach 6980, 76128 Karlsruhe,
Germany}
}
\maketitle
\begin{abstract}
We review recent results in heavy quark hadron
spectroscopy at the Tevatron.  With increasing data 
samples, the Tevatron experiments start to uncover
information on the spectroscopy of $b$-hadrons. Most important are the first
observations of the narrow \BsSS as well as \Sigmab,
\SigmabS and \Xibminus baryons. In addition we present
updated results on the narrow \BSS and \Bc mesons. 
\end{abstract}
%
%
\section{Introduction}
Heavy mesons consisting of a light quark and a heavy
anti-quark form an interesting laboratory for studying
QCD, the theory of strong interaction. 
They are a close analogue to the  hydrogen atom and
play a similar role for the study of QCD as hydrogen does 
for QED. The heavy anti-quark ($\overline{b}$ or
$\overline{c}$) takes the role of
the source of a static color potential, in which the light quark
($u$, $d$ or $s$) is located. Similarly, the heavy quark baryons
with a single heavy quark can in first order be viewed in the same
picture, only having a light diquark in the static color field
of the heavy quark. If the diquark picture isn't correct, then
one would arrive at an object similar to the helium atom with a heavy
quark generating the potential in which two light quarks are
located. Special case is the \Bc meson, which is the only
one composed by two distinct heavy quarks. The interplay of the two
heavy quarks, which decay through the weak interaction, is
important for our understanding of decays of the heavy quark
hadrons.

Heavy quark hadrons can be used to test QCD in regions
where perturbation calculations cannot be
used and many different approximations to solve the QCD have
been developed. Just a few examples of them are heavy quark
effective theory, non-relativistic and relativistic
potential models or lattice QCD. While a large amount of
information for $c$-hadrons exist\cite{Yao:2006px}, 
the spectroscopy of $b$-hadrons  was almost unknown up to recently. In this
paper we review recent measurements in the sector of heavy quark
hadrons by the CDF and D\O{} experiments at the Tevatron
collider. It is currently the only place where information about
 excited $b$-mesons and $b$-baryons can be obtained.
Charge-conjugate modes are implied
throughout this paper unless otherwise stated.

\section{Mass measurement of the \Bc meson}

Up to recently the \Bc meson was observed only in its 
semileptonic decay modes\cite{Abe:1998fb, Abulencia:2006zu}.
While  semileptonic decay modes have in general large
branching fractions,  the precision
of the mass measurement is rather limited due to the
undetected neutrino.  With increasing amount of 
data at the Tevatron, search in the fully reconstructed decay
modes becomes feasible.  The decay mode in which the \Bc search is done is
\BcJpsipi. The CDF collaboration obtained evidence for \BcJpsipi
decay\cite{Abulencia:2005usa} using 360 \invpb of data.

\begin{figure}[Htb]
  \begin{center}
    \includegraphics[width=7.0cm]{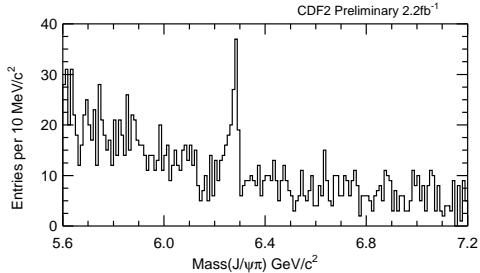}
    \caption{\it Invariant mass distribution of the \BcJpsipi
candidates observed by the CDF experiment.}
    \label{fig:BcMass}
  \end{center}
\end{figure}

This measurement\cite{BcPublic} of the 
\BcJpsipi decay is based on the data
selection developed on the high statistics \BplusJpsiK
decay. Its main feature is the huge background suppression
at a high signal efficiency. After the final selection we observe
around 19700 \BplusJpsiK signal events on $2.2$ \invfb of
data. Application of the same selection on $\Jpsi\pi^+$
sample yields to the invariant mass distribution shown in Fig.
\ref{fig:BcMass}.  A clear signal at a mass around $6270$ \mevcc is visible. 
To extract the mass and the number of \Bc signal candidates 
an unbinned maximum likelihood fit is used.
The signal is described by a Gaussian and the background by
a empirical function. The fit returns $87\pm 13$ signal events 
with a \Bc mass of  $6274.1 \pm 3.2 \stat \pm 2.6 \sys$
\mevcc. The statistical significance exceeds $8\sigma$. 
The measurement is compatible with existing predictions (see Ref.\cite{BcPublic}),
with an experimental uncertainty smaller than theory
uncertainties.
%

\section{Orbitally excited heavy quark mesons}

The bound states of a heavy $b$ anti-quark with a light $u$ or
$d$ quark are generically referred to as $B$ mesons. The
states with a light $s$ quark are analogous and are
referred to as a $\Bs$. The ground states with $J^P=0^0$ and
$J^P=1^-$ are well established\cite{Yao:2006px}, but
spectroscopy of the excited states has not been well
studied. The first excited state of the $B$ ($\Bs$) meson is
predicted to occur when a light quark has an orbital angular
momentum of $L=1$. Those states are collectively known as \BSS (\BsSS). 
Combining the spin of the light quark with its orbital momentum
yields two isodoublets with a total spin of light quark 
$J_l = 1/2$ and $J_l = 3/2$. The doublet $J_l = 1/2$
contains two states, $B^{*0}_0$ with total spin $J = 0$ and
$B_1^0$ with $J=1$. The members of the doublet with $J_l =
3/2$  are $\BSSone$ with $J = 1$ and $\BSStwo$ with $J=2$.
The $J_l = 1/2$ states decay to $B^{(*)}\pi$ 
via an $S$-wave transition.  Consequently, these states are 
expected to be very broad and difficult to observe at the
Tevatron. The $J_l = 3/2$ states decay to $B^{(*)+} \pi^-$ via a 
$D$-wave transition and are expected to be narrow. 
The decay $\BSSone \to B^+\pi^-$ is 
forbidden by angular momentum and parity conservation, while 
both $\BSStwo \to B^+\pi^-$ and $\BSStwo \to B^{*+}\pi^-$ decays are
allowed. The \BsSS system has the same structure, except of the $\pi^-$
changed to a $K^-$ in the decay. The decay of \BsSS to
$\Bs\pi^0$ is forbidden by isospin conservation.

\begin{figure}[Htb]
  \begin{center}
    \includegraphics[width=5cm]{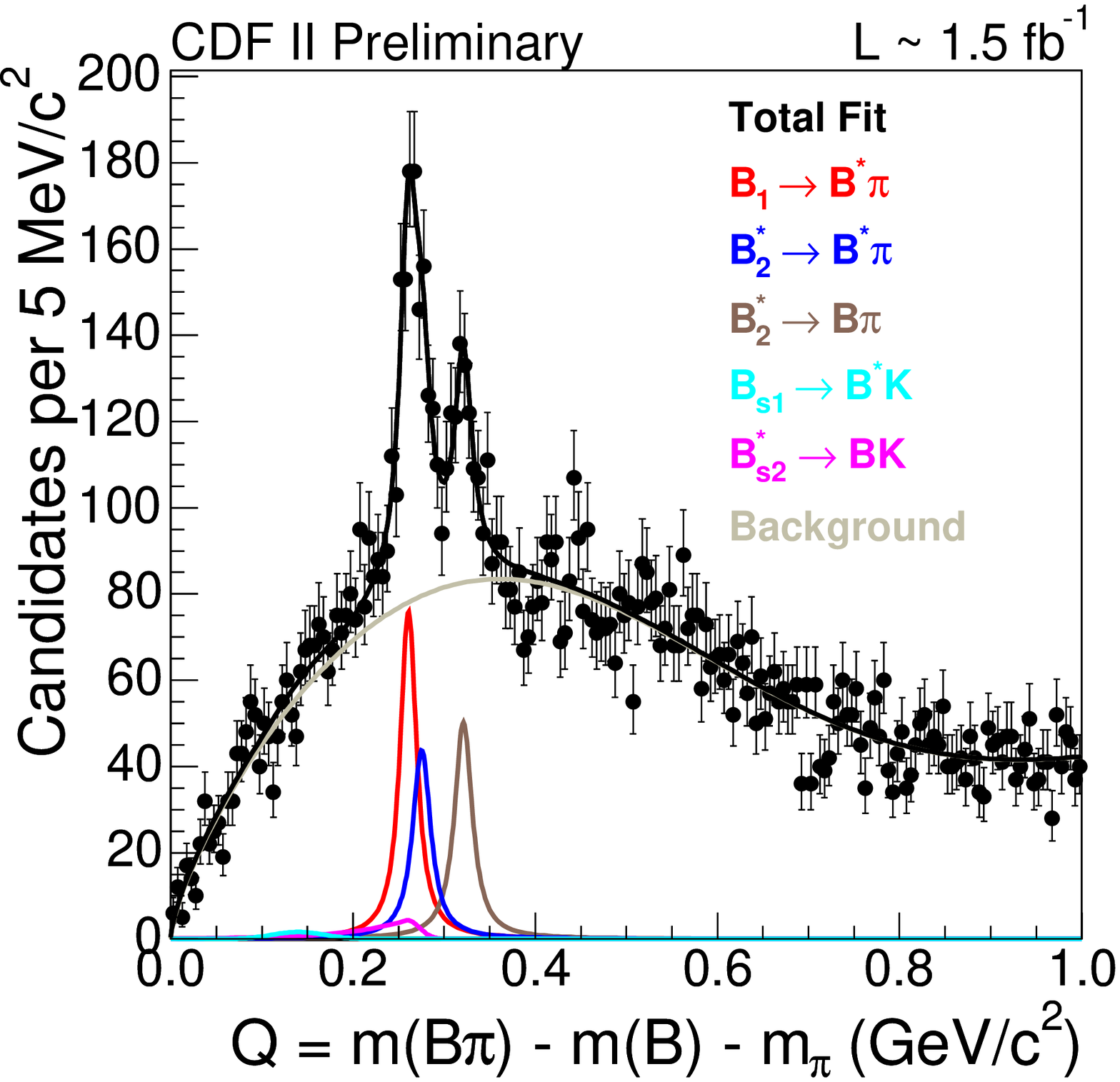}
    \includegraphics[width=5cm]{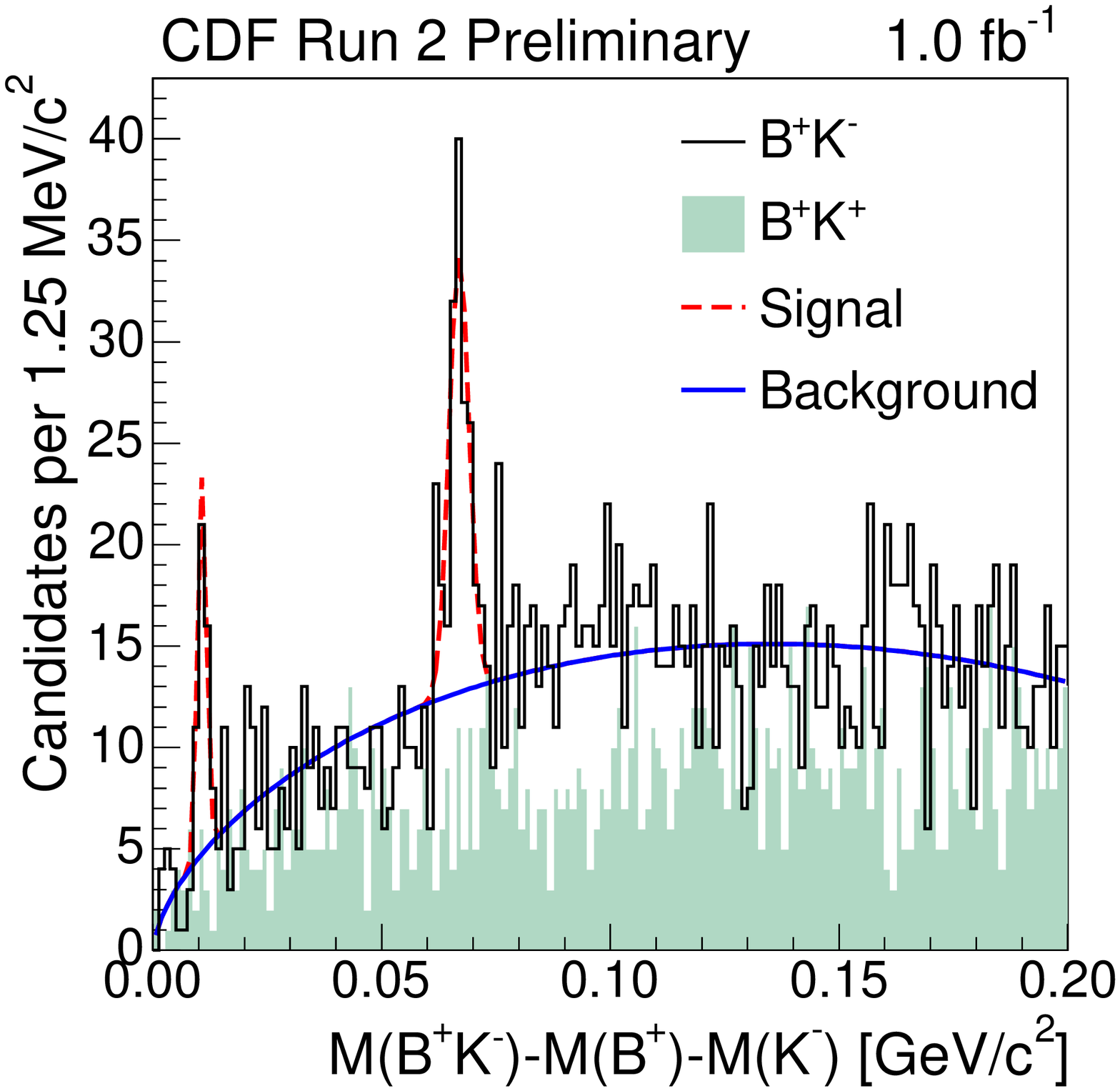}
    \caption{\it Invariant mass difference distribution for
$\Bplus \pi^-$ (left) and $\Bplus K^-$ (right) combinations
observed by the CDF experiment.}
    \label{fig:CDFBSS}
  \end{center}
\end{figure}

\begin{figure}[Htb]
  \begin{center}
    \includegraphics[width=5.0cm]{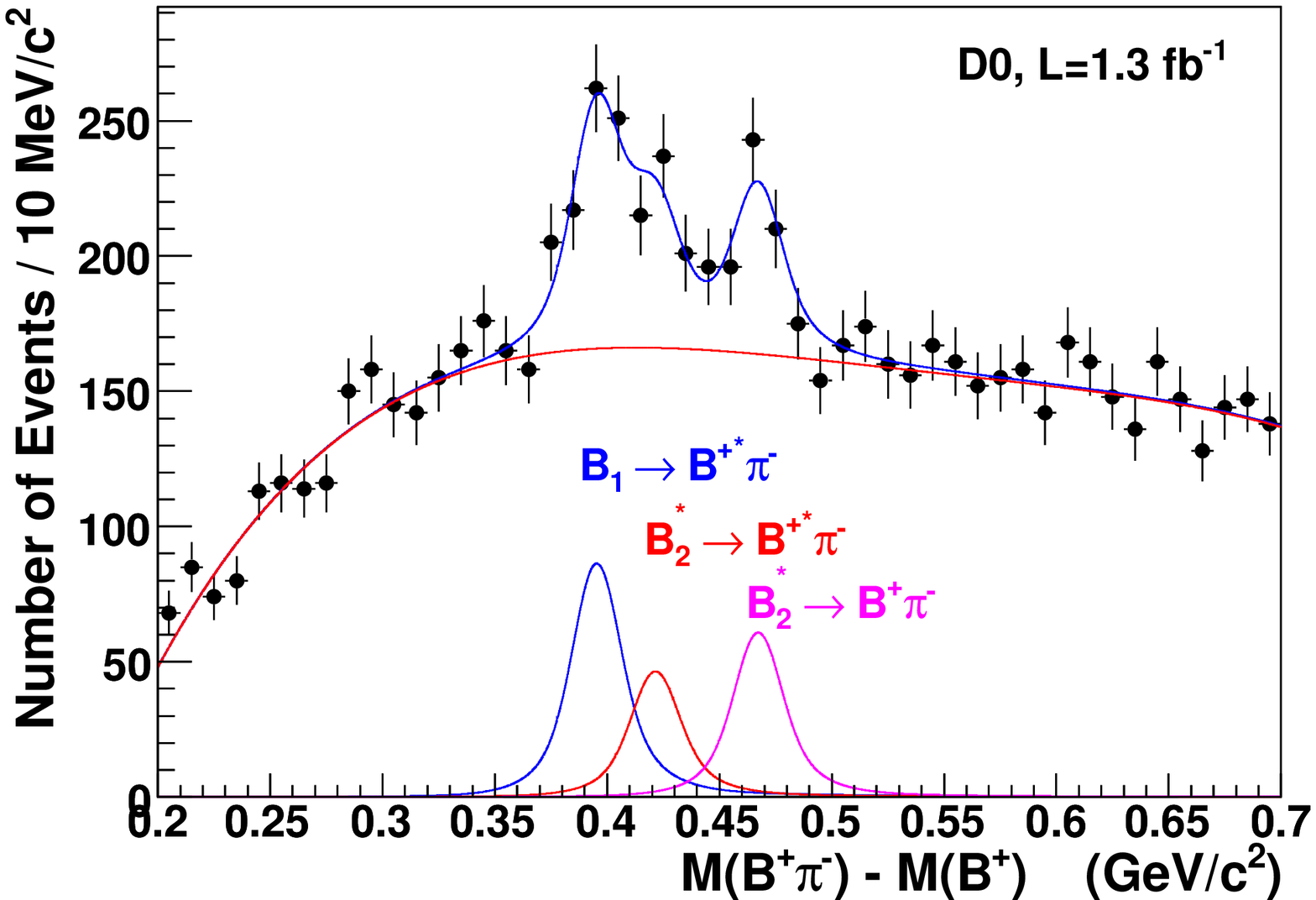}
    \includegraphics[width=5.0cm]{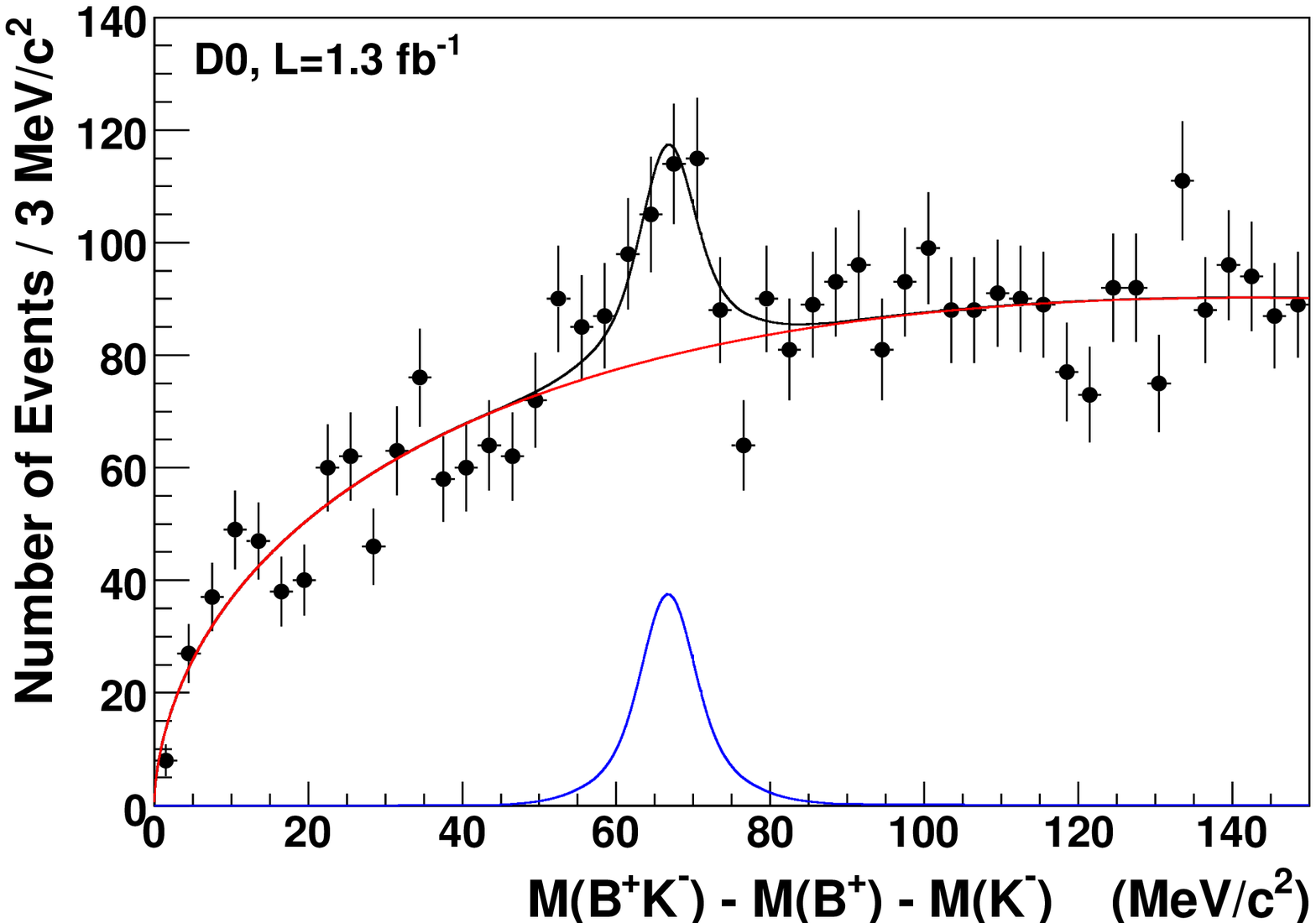}
    \caption{\it Invariant mass difference distribution for
$\Bplus \pi^-$ (left) and $\Bplus K^-$ (right) combinations
observed by the D\O{} experiment.}
    \label{fig:D0BSS}
  \end{center}
\end{figure}

Both Tevatron experiments perform studies of the narrow
\BSS\cite{BSS} and \BsSS\cite{BsSS} states in the $\Bplus
\pi^-$ and $\Bplus K^-$ final
states. The decays to \BS are included implicitly as \BS decays
to  $\Bplus\gamma$ with $\gamma$ undetected in both
experiments. The missing $\gamma$ will shift the reconstructed
mass by the mass difference between \BS and \Bplus. The \Bplus
is reconstructed in the $\Jpsi K^+$ final state by both
experiments. In addition, the $\DObar\pi^+$ mode  is used by
 CDF in the \BsSS search and the $\DObar3\pi$ mode is added to the
previous two for \BSS studies. 
The invariant mass difference of $\Bplus \pi^-$ and
$\Bplus K^-$ combinations obtained by CDF are shown in 
Fig.~\ref{fig:CDFBSS} and by D\O{} in Fig.~\ref{fig:D0BSS}.
For the first time experiments are able to observe the two
\BSS states as two separate peaks.
The measured masses are
listed in Table \ref{tab:excitedMesons}. 
In addition to the masses, the CDF experiment
measures for the first time also the width of the \BSStwo state
to be
$\Gamma(\BSStwo)=22.1^{+3.6}_{-3.1}\stat^{+3.5}_{-2.6}\sys$ \mev.
 Both experiments observe for the first time
the \BsSStwo state with a statistical significance larger than
$5\sigma$. The CDF experiment observes in addition
the \BsSSone state, which wasn't seen before, with a statistical
significance of more than $5\sigma$. 

\begin{figure}[Htb]
  \begin{center}
    \includegraphics[width=5.0cm]{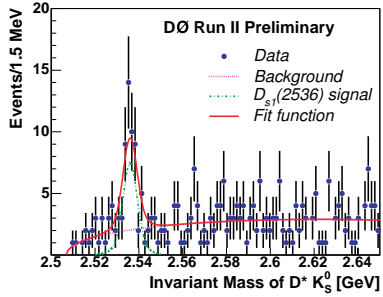}
     \caption{\it Invariant mass difference distribution for
$\DSplus K_s^0$ combinations observed by the D\O{}
experiment.}
    \label{fig:D0BsSS}
  \end{center}
\end{figure}
The D\O{} experiment performs also a mass measurement
of the \DsSSone state\cite{D0DsSSPublic}. The measurement is done in the
context of extraction of the  branching fraction
of the decay \BsDsonemunu.  In Fig.~\ref{fig:D0BsSS} the
invariant mass distribution of the $\DSplus K_s^0$
combinations coming from semileptonic \Bs decays is shown.
A very clean signal is obtained, which allows for a precise mass
measurement.
 For completeness, the branching fraction of the decay
\BsDsonemunu is measured to be $\mathcal{B}(\BsDsonemunu) =
        (0.86\pm0.16\stat\pm0.13\sys\pm0.09\mathrm{(ext)})\%$.
\begin{table}[t]
\centering
\caption{ \it Masses of the orbitally excited heavy quark
mesons. All values are in \mevcc with first uncertainty
being statistical and second systematical. }
\vskip 0.1 in
\tabcolsep=4mm
\renewcommand{\arraystretch}{1.3}
\begin{tabular}{|l|c|c|} \hline
 state & CDF & D\O \\
\hline \hline
\BSSone & $5725.3^{+1.6}_{-2.1}\ ^{+0.8}_{-1.1}$ & $5720.6\pm 2.4\pm 1.4$   \\
\BSStwo & $5739.9^{+1.7}_{-1.8}\ ^{+0.5}_{-0.6}$ & $5746.8\pm 2.4\pm 1.7$  \\
\BsSSone & $5829.4\pm 0.2 \pm 0.6$  & -  \\
\BsSStwo & $5839.0\pm 0.4 \pm 0.5$ & $5839.6\pm 1.1 \pm 0.7$ \\
\DsSSone & -  & $2535.7\pm 0.6\pm 0.5$ \\
\hline
\end{tabular}
\label{tab:excitedMesons}
\renewcommand{\arraystretch}{1.0}
\end{table}

\section{Observation of \Sigmab and \SigmabS baryons}

With increasing data samples collected at the Tevatron
accelerator,  searches for yet unobserved $b$-baryons begin to be
feasible. The first of such searches was performed by the CDF experiment,
which searched for the $\Sigma_b^\pm$ baryon and its spin
excited partner $\Sigma_b^{*\pm}$\cite{Pursley:2007zz}. 
A general theoretical expectations are the mass difference
$M(\Sigma_b)-M(\Lambda_b^0)-M(\pi)\,=\,40\,$--$\,70\,\mathrm{MeV}/c^2$
with $M(\Sigma_b^*)-M(\Sigma_b)\,=\,10\,$--$\,40\,\mathrm{MeV}/c^2$.
A small difference on the level of $5$ $\mathrm{MeV}/c^2$ is
expected between the masses of $\Sigma_b^+$ and $\Sigma_b^-$.
Both the $\Sigma_b$ and the $\Sigma_b^*$ are expected to be narrow
with a natural width of around $8$ and $15$ $\mathrm{MeV}/c^2$
with $\Lambda_b^0 \pi$ being the dominant decay mode.

The CDF search is based on $1$ \invfb of data using
fully reconstructed $\Lambda_b^0$ baryons. The $\Lambda_b^0$ is reconstructed in
the $\Lambda_c^+\pi^-$ decay mode with
$\Lambda_c^+\,\rightarrow\,pK^-\pi^+$. In total around $3200$
$\Lambda_b$ signal events are reconstructed. In the sample
used for the $\Sigma_b^\pm$ search $90$ \% of events are
$\Lambda_b^0$ baryons. The search is performed for the charged
$\Sigma_b^\pm$'s only, as the neutral one decays by emission of $\pi^0$,
which is extremely difficult to detect at the CDF experiment.

\begin{figure}[Htb]
  \begin{center}
    \includegraphics[width=6.5cm]{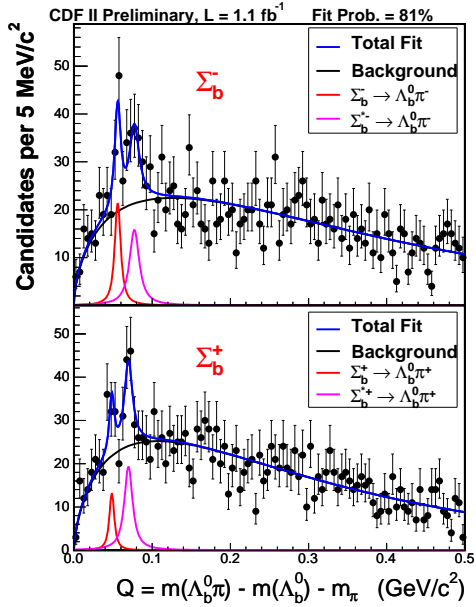}
    \caption{\it Projection of the fit result of the
$\Sigma_b^\pm$
invariant mass difference distribution. The points with
error bars represent the data. The blue line corresponds to the result of
the fit, the background is shown by the black line while the
signals are represented by the
red and magenta curves.}
    \label{fig:Sigmab}
  \end{center}
\end{figure}

The selected $\Lambda_b^0$ candidates are then combined with
charged pions to form $\Sigma_b^\pm$ candidates. After choosing 
the selection of candidates, the background is estimated while
keeping the signal region blinded. The background consists of
three basic components, which are combinatorial background,
$\Lambda_b^0$ hadronization and hadronization of
mis-reconstructed $B$ mesons. Relative fractions of these
components are taken from the fit of the $\Lambda_b^0$ invariant mass
distribution. The shape of the combinatorial background is
determined using the upper sideband of the $\Lambda_b^0$
invariant mass distribution. For the hadronization of mis-reconstructed $B$
mesons, the fully reconstructed $B^0\,\rightarrow\,D^-\pi^+$ in the data are used. The shape of the
largest component, $\Lambda_b^0$ hadronization, is determined
using a \textsc{pythia} Monte Carlo sample. The observed
invariant mass difference distribution is shown in Fig.
\ref{fig:Sigmab}.
To extract the signal yields and positions of the peaks,
an unbinned maximum likelihood fit is performed. The data are
described by a previously determined background shape
together with Breit-Wigner functions
convoluted with a resolution function for each peak. Due to
the low statistics, difference $M(\Sigma_b^{*+})\,-\,M(\Sigma_b^{+})$
is constrained to be the same as
$M(\Sigma_b^{*-})\,-\,M(\Sigma_b^{-})$.
The values obtained in the fit are summarized in Table
\ref{tab:sigmab} and the fit projection is shown in Fig.
\ref{fig:Sigmab}.
\begin{table}[tb]
\caption{\it Result of the fit to the $\Sigma_b$
invariant mass difference distribution. } 
\tabcolsep=4mm
\begin{center}
\renewcommand{\arraystretch}{1.3}
\begin{tabular}{|l|c|} \hline
 Parameter & Value \\ \hline \hline
$Q(\Sigma_b^+)$ ($\mathrm{MeV}/c^2$) & $48.5^{+2.0}_{-2.2}\ ^{+0.2}_{-0.3}$ \\
$Q(\Sigma_b^-)$ ($\mathrm{MeV}/c^2$) & $55.9\pm1.0\pm0.2$ \\
$M(\Sigma_b^{*})-M(\Sigma_b)$ ($\mathrm{MeV}/c^2$) & $21.2^{+2.0}_{-1.9}\ ^{+0.4}_{-0.3}$ \\ \hline
$\Sigma_b^+$ events & $32^{+13}_{-12}\ ^{+5}_{-3}$ \\
$\Sigma_b^-$ events & $59^{+15}_{-14}\ ^{+9}_{-4}$ \\
$\Sigma_b^{*+}$ events & $77^{+17}_{-16}\ ^{+10}_{-6}$ \\
$\Sigma_b^{*-}$ events & $69^{+18}_{-17}\ ^{+16}_{-5}$\\ \hline
\end{tabular}
\renewcommand{\arraystretch}{1.0}
\end{center}
\label{tab:sigmab}
\end{table}

To estimate the significance of the observed signal, the fit is
repeated with an alternative hypothesis and the difference in the
likelihoods is used. Three different alternative hypotheses were
examined, namely the null hypothesis, using only two peaks instead of
four and leaving each single peak separately out of the fit. As a
result we conclude that the null hypothesis can be excluded by
more than five standard deviations. The fit 
favors four peaks against two and except of the $\Sigma_b^+$
peak, each peak has a significance above three standard
deviations.

\section{Observation of the \Xibminus baryon}

The latest state observed by the Tevatron experiments is
the \Xibminus baryon\cite{Xib}, a state with  quark content
$dsb$. The mass of the \Xibminus is expected to be around
$5.8$ \gevcc. The decay is dominated by the weak decay of the
$b$ quark. The LEP experiments observed excess in  $\Xi^-
l^-\nu_l X$ events, which was attributed to the \Xib baryon and
the lifetime $\tau=1.39^{+0.34}_{-0.28}$ \ps was deduced\cite{Yao:2006px}.
Suitable decay modes for the search at the Tevatron are
\XibminusJpsiXi, which can be used by both CDF and D\O{} and
the \XibXicpi, \XibDLambda, \XibLambdacKpi decay modes
accessible at the CDF experiment. The presented search
uses the decay mode  \XibminusJpsiXi which has the advantage
of a \Jpsi in the final state leading to clean trigger signature.
A disadvantage of the used decay mode is that only the \Xibminus
is accessible as the \Xibzero contains $\pi^0$ in the decay chain.

A complication in the study of the \Xibminus state comes from having a
$\Xi$ in the final state, which decays through the weak interaction to
$\Lambda$ and $\pi$ with a subsequent decay of
$\Lambda\rightarrow p\pi$. As both $\Xi$ and $\Lambda$ have
long lifetime, their decay vertices are significantly
displaced from the production point. This requires a special
treatment of the track reconstruction comparing to the usual
tracks used in $b$-hadron studies. In addition the $\Xi^-$ is
charged and travels several centimeters in the magnetic field
which adds to the complexity of the analysis as the bending of
the $\Xi^-$ is significant. On the other hand there is
a possibility to gain in precision of the secondary vertex
resolution by tracking $\Xi^-$ in the silicon detector close
to the interaction region. The CDF experiment chosed this approach, 
leading to  improvements in the precision of the $\Xi^-$
impact parameter measurement as well as in 
determination of the \Xibminus secondary vertex position.

\begin{figure}[Htb]
  \begin{center}
    \includegraphics[width=5cm]{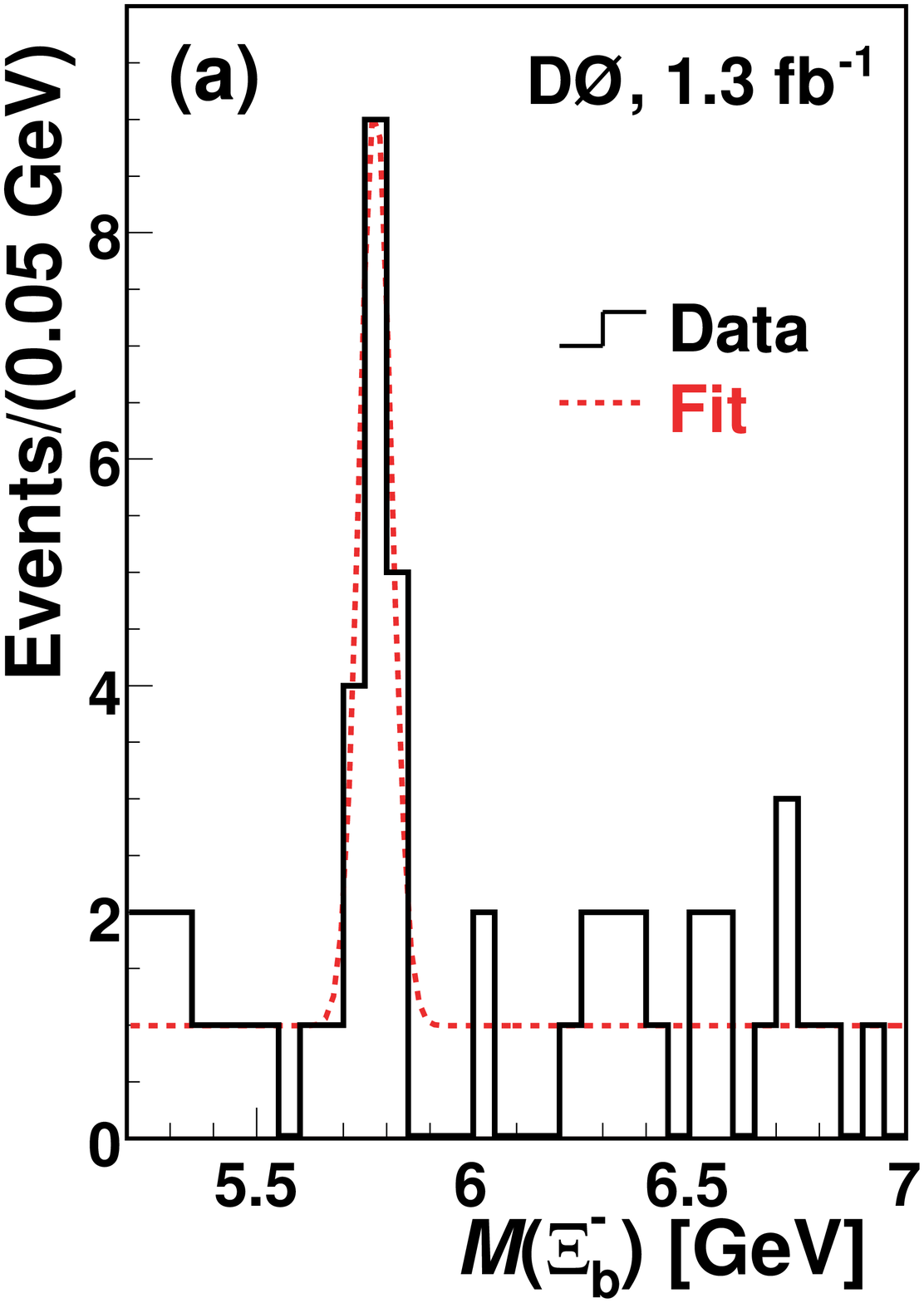}
    \includegraphics[width=5cm]{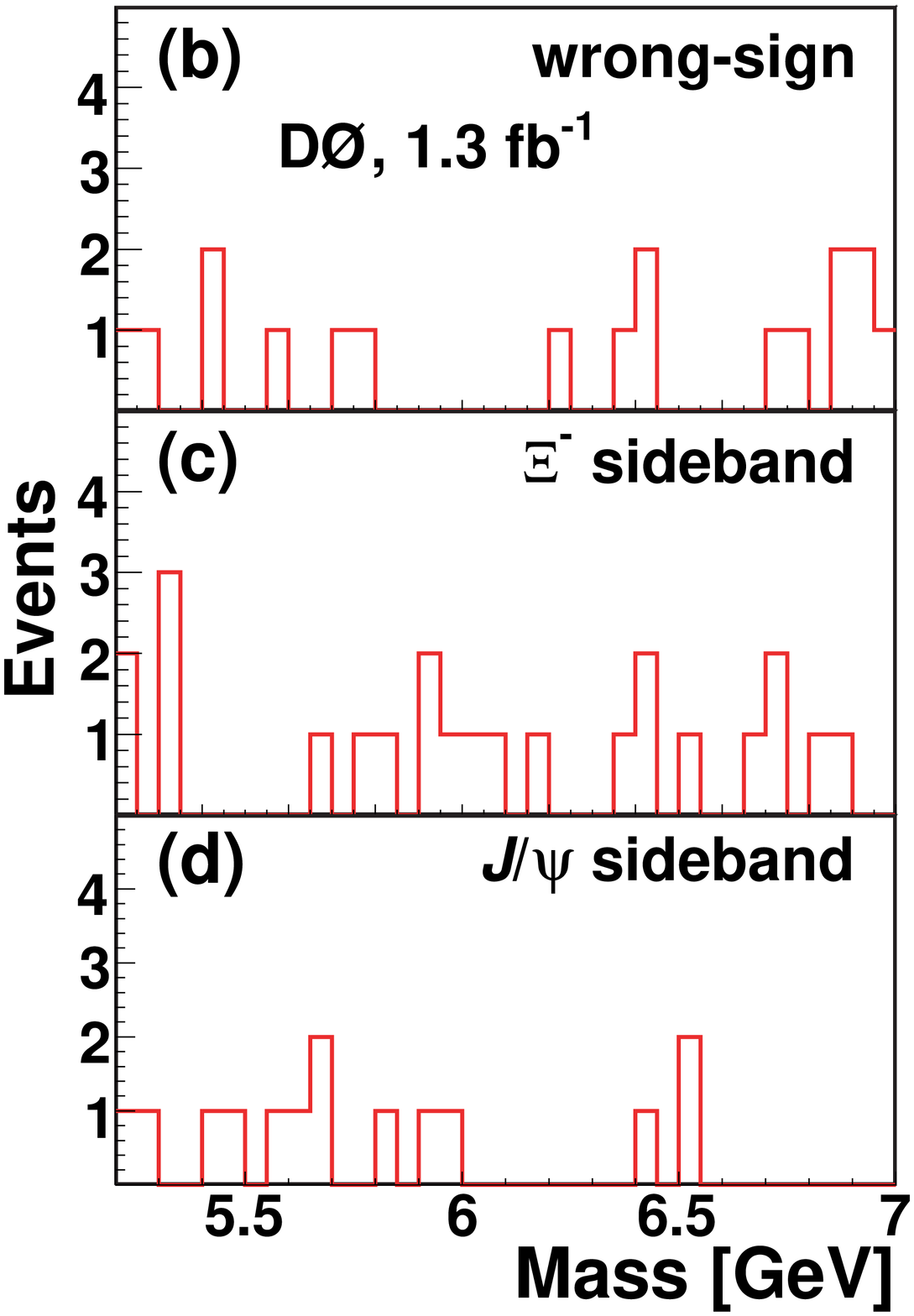}
    \caption{\it  Invariant mass distribution of the
\Xibminus candidates observed by the D\O{} experiment (left) and
invariant mass of the various background samples (right).}
    \label{fig:D0Xib}
  \end{center}
\end{figure}

\begin{figure}[Htb]
  \begin{center}
    \includegraphics[width=8cm]{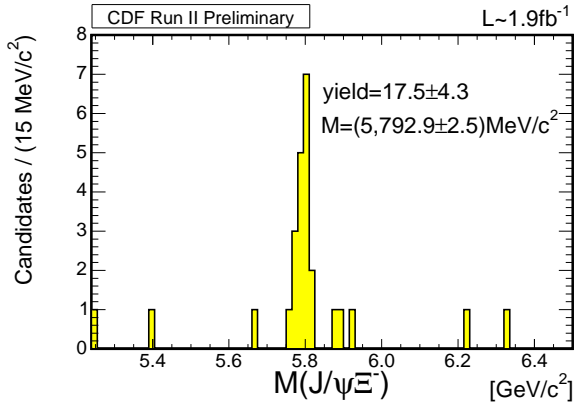}
    \caption{\it Invariant mass distribution of the
\Xibminus candidates observed by the CDF experiment.}
    \label{fig:CDFXib}
  \end{center}
\end{figure}

Both experiments use the momenta of the \Xibminus candidate and its
daughters, vertex quality along with the \Xibminus decay vertex
displacement to select final candidates. The D\O{} experiment
develops the selection based on the signal from simulated events
and background from wrong-sign data.
The invariant mass distribution of selected candidates is
shown in Fig. \ref{fig:D0Xib}. The CDF collaboration uses
a data only approach. As the $\Xi^-$ is tracked in the
silicon detector, one can treat it as an ordinary track. In
such an approach, the decay \XibminusJpsiXi is similar to the decay
\BplusJpsiK. This allows to reuse the selection developed on
\BplusJpsiK decays for the \Bc search. The observed invariant
mass distribution is shown in Fig. \ref{fig:CDFXib}. In both
experiments a clear signal with a mass slightly below $5.8$
\gevcc is visible. 

To extract the mass and number of signal events, both
experiments perform an unbinned maximum likelihood fit. They
obtain the number of signal events
$N_s=14.8\pm4.3\stat^{+1.9}_{-0.4}\sys$ (D\O) and
$N_s=17.5\pm4.3\stat$ (CDF). The statistical significance 
of the signal is $5.2\sigma$ and $7.7\sigma$ for the D\O{} and CDF 
experiment respectively.  
The measured masses are $5774\pm11\stat\pm15\sys$ \mevcc (D\O) and
$5792.9\pm2.5\stat\pm1.7\sys$ \mevcc (CDF) are in good agreement
between the two experiments.

Several cross checks were done by both experiments to
strengthen the interpretation of the observed signal as
a \Xibminus state. The D\O{} experiment made a detailed
examination of the wrong-sign combinations together with the
$\Xi^-$ and \Jpsi sideband events with no signal observed in
any of these (see Fig. \ref{fig:D0Xib}). 
In addition the proper decay length distribution
from data was compared to the one expected for a typical
weakly decaying $b$-hadron and  good consistency was
observed.  The CDF experiment used its unique
opportunity to trigger on  events with displaced vertices and
searched also for the \XibminusXicpi decay mode. Also here 
evidence for the signal is seen with the mass at same position
as in \XibminusJpsiXi. Thus one can
conclude that the observed signal is due to the \XibJpsiXi
decay.

\section{Conclusions}

Heavy quark states provide an interesting laboratory for
testing various approaches to the non-perturbative regime of
QCD.  The Tevatron experiments have made large effort to
improve our knowledge of the $b$-hadrons. Roughly one and
half year ago only few $b$-mesons were known. The \Bc meson
was seen only in a semileptonic decay mode with a large 
uncertainty on the mass measurement. Orbitally excited mesons were
not observed (\BsSS) or could not be seen as distinct
peaks (\BSS). In the $b$-baryon sector, only \Lambdab was
directly observed with little information on \Xib obtained
by the LEP experiments from the excess in $\Xi^- l^-\nu_l X$
events.

Since then the effort of the CDF and D\O{} collaborations
provided new important data. It started with the observation of
fully reconstructed \BcJpsipi decay at CDF, which allowed for
a precise mass measurement. 
Both experiments contributed to 
studies of the orbitally excited $B$ and \Bs mesons. From those
studies both $J_l=3/2$ states of the \BsSS have been observed for
the first time. Also the $J_l=3/2$ states of the \BSS have
been for the first time seen as two separate peaks. 
On the side of $b$-baryons,  \Sigmab and \SigmabS as well
as \Xibminus were observed starting a new era in the study of 
$b$-baryons. 

To conclude, due to the effort of the Tevatron experiments our
knowledge of the $b$-hadrons was increased considerably,
but lot of room to improve our knowledge on
the properties of already observed hadrons still exists. On the side of
unobserved hadrons, the next
focus should be on the $\eta_b$ search, the last
unobserved meson containing a $b$ quark. Also an observation of the
\Xibzero, which should be possible at  CDF, would
strengthen the interpretation of the signal
attributed to \Xibminus. Last, an observation of the \Omegab
would be a nice completion of the Tevatron program on 
spectroscopy of $b$-hadrons. With more data coming and
the well understood detector we believe that at least some of 
those searches will be successful.

\section*{Acknowledgments}
The author would like to thank all his colleagues from the CDF
and D\O{} experiments for performing the studies presented here
as well as for their help in the preparation of the talk and this
paper.


%
\end{document}